\documentclass[prd,superscriptaddress,showpacs,twocolumn]{revtex4}

\usepackage{amsmath}
\usepackage{amssymb}
\usepackage{slashed}
\usepackage[dvips]{graphicx}

\newcommand{\bra}[1]{\langle#1\vert}
\newcommand{\ket}[1]{\vert#1\rangle}
\newcommand{\Tr}{\mathop{\rm Tr}\nolimits}
\newcommand{\I}{\ensuremath{\mathrm{i}}}
\newcommand{\e}{\ensuremath{\mathrm{e}}}
\renewcommand{\Re}{\mathop{\rm Re}\nolimits}
\newcommand{\eL}{\mathcal{L}}
\newcommand{\D}{\ensuremath{\mathrm{D}}}
\renewcommand{\d}{\ensuremath{\mathrm{d}}}
\newcommand{\cc}{\ensuremath{\mathrm{c.c.}}}

\begin{document}

\title{Dynamical breakdown of Abelian gauge chiral symmetry by strong Yukawa interactions}

\author{Petr Bene\v s}
\email{benes@ujf.cas.cz}
\affiliation{Department of Theoretical Physics, Nuclear Physics Institute, \v Re\v z (Prague), Czech Republic}
\affiliation{Faculty of Mathematics and Physics, Charles University, Prague, Czech Republic}

\author{Tom\'a\v s Brauner}
\email{brauner@ujf.cas.cz}
\affiliation{Department of Theoretical Physics, Nuclear Physics Institute, \v Re\v z (Prague), Czech Republic}
\affiliation{Faculty of Mathematics and Physics, Charles University, Prague, Czech Republic}

\author{Ji\v r\'{\i} Ho\v sek}
\email{hosek@ujf.cas.cz}
\affiliation{Department of Theoretical Physics, Nuclear Physics Institute, \v Re\v z (Prague), Czech Republic}

\begin{abstract}
We consider a model with anomaly-free Abelian gauge axial-vector symmetry,
which is intended to mimic the standard electroweak gauge chiral $SU(2)_L\times
U(1)_Y$ theory. Within this model we demonstrate: (1) Strong Yukawa
interactions between massless fermion fields and a massive scalar field
carrying the axial charge generate dynamically the fermion and boson proper
self-energies, which are ultraviolet-finite and chirally noninvariant. (2)
Solutions of the underlying Schwinger-Dyson equations found numerically exhibit
a huge amplification of the fermion mass ratios as a response to mild changes
of the ratios of the Yukawa couplings. (3) The `would-be' Nambu-Goldstone boson
is a composite of both the fermion and scalar fields, and it gives rise to the
mass of the axial-vector gauge boson. (4) Spontaneous breakdown of the gauge
symmetry further manifests by mass splitting of the complex scalar and by new
symmetry-breaking vertices, generated at one loop. In particular, we work out
in detail the cubic vertex of the Abelian gauge boson.
\end{abstract}

\pacs{11.30.Qc}
\maketitle

\section{Introduction}

Use of the general principle of spontaneous breakdown of continuous symmetry
for the $SU(2)_L\times U(1)_Y$ invariant field theory description of
electroweak phenomena is a necessity. Hard symmetry-breaking mass terms of both
gauge and fermion fields ruin, either directly or by virtue of the loops,
decent high-energy behavior of certain scattering amplitudes.

In this respect the standard `Higgs' realization of spontaneous breakdown of
the electroweak symmetry deserves admiration. At the expense of introducing a
sector with a condensing scalar doublet it defines the operational
(renormalizable, weakly coupled) description of virtually all electroweak
phenomena so far explored.

Because of theoretical drawbacks of the Higgs mechanism there are alternatives
and/or generalizations thereof \cite{Eidelman:2004wyTC} (and references
therein). In the near future the Large Hadron Collider at CERN will harshly
test all of them. The aim of this paper is to add to the already existing list
yet another realization of spontaneous breakdown of chiral and gauge symmetry.
For clarity we illustrate it on physically and technically transparent Abelian
prototype. Comparison with the standard Abelian Higgs mechanism can easily be
made at any stage by heart.

The basic idea \cite{Brauner:2005hw} is simple but subtle: As in the standard
Higgs realization we employ the complex scalar field but with an ordinary mass
term of a complex scalar. Hence, in accordance with common wisdom there will be
no spontaneous symmetry breakdown in the scalar sector itself. Subtle is the
nonperturbative self-consistence: The chiral symmetry breaking fermion proper
self-energy is both caused and causes the symmetry-breaking scalar proper
self-energy, both solely due to a strong Yukawa interaction.

The subsequent question of whether a massless gauge field introduced by gauging
the $U(1)$ axial symmetry describes a massless or a massive spin-1 particle is
dynamical and it was in general answered by Schwinger \cite{Schwinger:1962tn}:
If, within a given dynamics, the gauge field polarization tensor develops a
massless pole, its residue approximates the gauge boson mass squared
\cite{Jackiw:1973tr}. Both in the Abelian Higgs model and in the model
presented below the massless pole in the gauge field polarization tensor is due
to the `would-be' Nambu-Goldstone (NG) boson of the underlying global symmetry.
While in the Higgs mechanism the NG boson is pre-prepared in the elementary
scalar field, in our case it is a composite object of both the fermion and the
scalar fields.

Finally, it is easy to realize that due to the dynamical generation of the
symmetry-breaking pieces in fermion and scalar field propagators there are new
effective symmetry-breaking vertices between the mass eigenstates of the
fields.

The paper is organised as follows: After introducing the model by its
Lagrangian we discuss the possibility of the dynamical spontaneous breaking of
the chiral symmetry and introduce the appropriate formalism. Then we explore
the consequences of the global symmetry: the existence of the `would-be' NG
boson and its interaction vertices with other particles. In this part we only
repeat the main results of our preceding paper \cite{Brauner:2005hw} so we do
not go into a detail. After this we present some consequences of the local
symmetry: the mass of the gauge boson, induced by the bilinear coupling with
the `would-be' NG boson, and the existence of the effective trilinear coupling
of the gauge bosons. In the end we present our numerical results, especially
the possibility of an arbitrarily high ratio of fermion masses with Yukawa
coupling constants being of the same order of magnitude, together with the
description of our numerical procedure.

\section{The Model}

In the full generality the model is defined by the Lagrangian
\begin{eqnarray}
    \eL & = & \bar\psi_1\I\slashed{\D}\psi_1+\bar\psi_2\I\slashed{\D}\psi_2
    +(\D_{\mu}\phi)^\dag(\D^{\mu}\phi) \nonumber \\
    &&
    {}-M^2\phi^{\dagger}\phi
     -\frac12\lambda(\phi^{\dagger}\phi)^2 -\frac14 F_{\mu\nu}F^{\mu\nu} \\
    && {}+ \eL_{\rm{Yukawa}} \nonumber \,,
    \label{lagrangian}
\end{eqnarray}
where the Yukawa interaction Lagrangian reads
\begin{equation}
\begin{split}
    \eL_{\rm{Yukawa}}
    & \> = \>  \hphantom{+\,}
    y_1(\bar\psi_{1L}\psi_{1R}\phi+\bar\psi_{1R}\psi_{1L}\phi^{\dagger})\\
    &  \hphantom{\> = \>} \, +\,
    y_2(\bar\psi_{2R}\psi_{2L}\phi+\bar\psi_{2L}\psi_{2R}\phi^{\dagger}) \,.
\end{split}
\end{equation}
The Lagrangian possesses various symmetries. First, it is invariant under the
\emph{global} $U(1)_{V_1}\times U(1)_{V_2}\times U(1)_A$ symmetry. The two
global vector $U(1)$ symmetries are generated by the fermion number operators
of the fermions 1 and 2 and represent separate conservation of the number of
both fermion species. The two naive independent axial $U(1)$ transformations of
the fermions 1 and 2 are related to each other by the Yukawa couplings to the
scalar field $\phi$. Second, the Lagrangian is invariant under the \emph{local}
(gauge) $U(1)_A$ transformations, acting on the fields as
\begin{eqnarray}
    \psi_j  & \rightarrow & \e^{ \I \theta(x) Q_j \gamma_5} \psi_j \,, \nonumber \\
    \phi    & \rightarrow & \e^{-2 \I \theta(x)}\phi \,, \\
    A_{\mu} & \rightarrow & A_{\mu} + \frac{1}{g}\partial_\mu \theta(x) \nonumber \,,
\end{eqnarray}
where the axial charges $Q_1$, $Q_2$ are defined as
\begin{equation}
\begin{split}
    Q_1 & \> = \>  +1  \,, \\
    Q_2 & \> = \>  -1  \,.
\end{split}
\end{equation}

Considering two fermion species with opposite axial charges provides the
simplest solution to an important theoretical constraint imposed on the model,
which is the absence of the axial anomaly. The Yukawa coupling constants $y_1$,
$y_2$ are arbitrary real numbers. For the sake of simplicity we do not consider
a more general Yukawa Lagrangian, allowing them to be complex.

\section{Global Symmetry}

\subsection{Self-energies \& masses}

In the first step we switch off the gauge interaction $(g=0)$ and demonstrate
the spontaneous breakdown of the global $U(1)_A$ symmetry by finding
symmetry-breaking proper self-energies in the fermion and scalar field
propagators. We follow the general self-consistent nonperturbative method
suggested by Nambu (and Jona-Lasinio) \cite{Nambu:1960tm,Nambu:1961tp}.

First of all, it is convenient to introduce the Nambu-like doublet field for
the scalar
\begin{equation}
\Phi=
\left(
\begin{array}{c}
\phi \\
\phi^{\dagger}
\end{array}
\right)
\end{equation}
and its matrix propagator
\begin{equation}
\begin{split}
\I D(x-y) & = \bra{0}T\{\Phi(x)\Phi^{\dagger}(y)\}\ket{0} \\
& =
\left(
\begin{array}{cc}
\langle\phi\phi^\dag\rangle & \langle\phi\phi\rangle \\
\langle\phi^\dag\phi^\dag\rangle & \langle\phi^\dag\phi\rangle
\end{array}
\right) \,.
\end{split}
\end{equation}
The point is that this formalism allows us to treat the symmetry-preserving
(the diagonal entries) as well as the symmetry-breaking (the off-diagonal
entries) scalar propagators on the same footing. Similarly in the fermion
sector, it is convenient to deal with the propagator
\begin{equation}
\I S(x-y)=\bra{0}T\{\psi(x)\bar\psi(y)\}\ket{0} \,,
\end{equation}
as it also contains both the symmetry-preserving (e.g.
$\langle\psi_L\bar\psi_L\rangle$) and the symmetry-breaking (e.g.
$\langle\psi_L\bar\psi_R\rangle$) fermion propagators.

However, although it would be possible, and even desirable, to deal with the
propagators in the full generality, we will adopt for the present purposes the
specific \emph{Ansatz}:
\begin{equation}
\begin{split}
S_{j}^{-1}(p) & \> = \> \slashed{p}-\Sigma_{j}(p^2) \,, \\
D^{-1}(p) & \> = \> \left(
\begin{array}{cc}
p^2-M^2 & -\Pi(p^2) \\
-\Pi^*(p^2) & p^2-M^2
\end{array}\right) \,.
\end{split}
\label{ansatz}
\end{equation}
Here the functions $\Sigma_j(p^2)$ and $\Pi(p^2)$ (in the following text often
denoted simply as $\Sigma_{j,p}$ and $\Pi_p$) are the one particle irreducible
(1PI) parts of the corresponding \emph{symmetry-breaking} propagators, i.e. the
proper self-energies. By this Ansatz we focus only on the symmetry-breaking
parts of the propagators and totally neglect any possible effects of
wave-function renormalization. For the purposes of the present paper this
approximation is sufficient, since we wish only to demonstrate spontaneous
symmetry breaking and not to make at this stage any phenomenological
predictions. In a more realistic version of the model aspiring to describe the
Nature and make reasonable physical predictions it would be, however, a must to
include systematically all radiative corrections.

Our general strategy in demonstrating the spontaneous breakdown of the $U(1)_A$
symmetry will be to search for the symmetry breaking parts of the propagators,
i.e. the self-energies $\Sigma_{j}$ and $\Pi$. The key observation is, however,
that at any finite order of perturbative expansion the $U(1)_A$ symmetry
remains preserved and the self-energies vanish. The spontaneous symmetry
breaking is a \emph{nonperturbative effect}. To treat it one has to employ some
nonperturbative technique. We have chosen to make use of the
\emph{Schwinger-Dyson equations}, which represent a formal summation of all
orders of perturbative expansion and as such they provide the desired
nonperturbative treatment.

\begin{figure}[t]
\begin{center}
\includegraphics{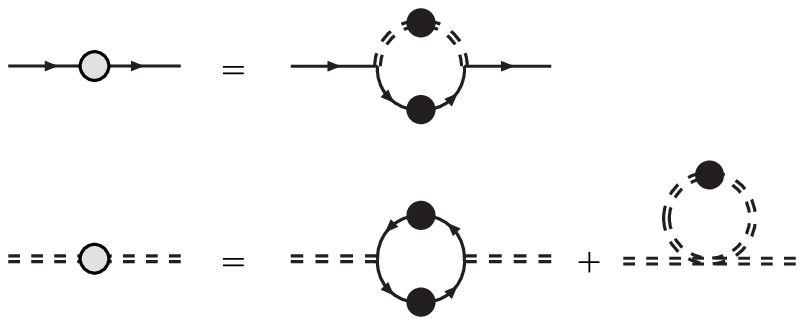}
\caption[]{The diagrammatical representation of Schwinger-Dyson equations (\ref{SD_model_equations}). The first line holds for
both $\psi_1$ and $\psi_2$. The grey blobs stand for the proper
self-energies, while the solid black blobs denote the full propagators. The double
dashed line is for the Nambu $\Phi$ doublet.}\label{sd-pic}
\end{center}
\end{figure}

The system of Schwinger-Dyson equations forms an infinite tower of equations
for all Green's functions of the theory. For practical calculation one usually
has to truncate it at some level. We truncated it at the level of three-point
Green's functions, which we approximate by the bare ones. The resulting system
of Schwinger-Dyson equations in terms of the unknown functions $\Sigma_{j}$ and
$\Pi$ is
\begin{widetext}
\begin{equation}
\begin{split}
\Sigma_{1,p}&=\I y_1^2\int\frac{\d^4k}{(2\pi)^4}\frac{\Sigma_{1,k}}{k^2-\Sigma_{1,k}^2}
\frac{\Pi_{k-p}}{\left[(k-p)^2-M^2\right]^2-|\Pi_{k-p}|^2} \,,\\
\Sigma_{2,p}&=\I y_2^2\int\frac{\d^4k}{(2\pi)^4}\frac{\Sigma_{2,k}}{k^2-\Sigma_{2,k}^2}
\frac{\Pi^*_{k-p}}{\left[(k-p)^2-M^2\right]^2-|\Pi_{k-p}|^2} \,,\\
\Pi_p&=-\sum_{j=1,2}2\I y_j^2\int\frac{\d^4k}{(2\pi)^4}\frac{\Sigma_{j,k}}{k^2-\Sigma_{j,k}^2}
\frac{\Sigma_{j,k-p}}{(k-p)^2-\Sigma_{j,k-p}^2}+\I\lambda\int\frac{\d^4k}{(2\pi)^4}
\frac{\Pi_k}{(k^2-M^2)^2-|\Pi_k|^2} \,.
\end{split}
\label{SD_model_equations}
\end{equation}
\end{widetext}
See also the diagrammatical representation of these Schwinger-Dyson equations
-- Fig. \ref{sd-pic}. If nonzero solutions $\Sigma_{j}$ and $\Pi$ exist, then
they should necessarily be ultraviolet(UV)-finite since the corresponding
counter terms are prohibited by symmetry.

Fermion masses $m_{j}$ are determined in terms of $\Sigma_{j}$ by solving
\begin{equation}
m_{j}^2 = \Sigma^2_{j}(p^2=m^2_{j}) \,.
\label{mass_fermions}
\end{equation}
By dimensional arguments the solutions must have the form
\begin{equation}
m_{j} = M f_{j}(y_1,y_2) \,,
\label{f-mass}
\end{equation}
to be compared with $m_{j}=(-2M^2/\lambda)^{1/2} y_{j}$ in the case of a
condensing $\phi$. This is the crucial point: while in the standard Higgs
mechanism the functions $f_j$ are simple linear functions, in the present model
we expect some more complicated functions, which should also be nonanalytic,
because we deal with nonperturbative effects, that can not be obtained at any
finite order of perturbative expansion in coupling constants $y_{j}$. Moreover
there is a possibility that a small change in $y_{j}$ (say, about one order of
magnitude) might produce much larger change in order of magnitude of fermion
masses.

Like in the fermionic sector, the scalar spectrum can be obtained as a pole of
the propagator:
\begin{equation}
M_{1,2}^2 = M^2 \pm  |\Pi(p^2=M^2_{1,2})| \,.
\label{mass_scalars}
\end{equation}
This is easily interpreted: As a result of spontaneous symmetry breaking there
appear two real scalar fields with \emph{different} masses $M_{1,2}$, instead
of one complex field with mass $M$.

Our numerical analysis confirms the existence of the nonzero solutions
$\Sigma_{j}$ and $\Pi$. At present we are, however, able to say nothing about
their uniqueness. Upon performing the Wick rotation we have found numerically
the real solutions $\Sigma_{j}$ and $\Pi$ with the following properties: First,
they vanish very fast at large $p^2$ (faster then any power of $p^2$). Second,
the solutions are found so far only for large values of the Yukawa coupling
constants $y_{j}$ (or, more precisely, for $y_1$ and $y_2$ not being
simultaneously small). Third, fermion mass ratio $m^2_1/m^2_2$ exhibits
tremendous amplification upon decent changes of $y_1/y_2$. For example, for
$y_1/y_2=77.4/88$ we get $m^2_1/m^2_2= 10^{-2}$. Obviously this is alluring. If
justified and understood analytically, this property of the solutions of the
model should not be called fine tuning. Finding the explicit form of the
functions $f_j(y_1,y_2)$ in Eq. (\ref{f-mass}) is our ultimate dream.

More details on the numerics and adopted approximations can be found in the
Sec. \ref{numerics}.

\subsection{Nambu-Goldstone boson}

With the gauge interaction still switched off we reveal in the second step the
Nambu-Goldstone (NG) excitation and compute its effective couplings with
fermion and boson fields. The analysis is standard:

The conservation of the axial-vector current
\begin{equation}
j^{\mu}_{A}(x) = \bar\psi_1\gamma^{\mu}\gamma_5\psi_1-\bar\psi_2\gamma^{\mu}\gamma_5\psi_2+
2\I\left[(\partial^{\mu}\phi)^{\dagger}\phi-\phi^{\dagger}\partial^{\mu}\phi\right]
\end{equation}
implies the axial-vector Ward identities for the proper vertex functions:
\begin{equation}
\begin{split}
q_{\mu}\Gamma^{\mu}_{\psi_j}(p+q,p) & \> = \> Q_j \left[ S_j^{-1}(p+q)\gamma_5+\gamma_5S_j^{-1}(p) \right] \,,\\
q_{\mu}\Gamma^{\mu}_{\phi}(p+q,p)   & \> = \> -2D^{-1}(p+q)\Xi+2\Xi D^{-1}(p) \,,
\end{split}
\label{Ward_identities}
\end{equation}
where the proper vertex functions $\Gamma^{\mu}_{\psi_j}$,
$\Gamma^{\mu}_{\phi}$ correspond to the Green's functions
\begin{equation}
\begin{split}
G_{\psi_j}^{\mu}(x,y,z) & \> = \> \bra{0}T\{j_A^{\mu}(x)\psi_j(y)\bar\psi_j(z)\}\ket{0} \,, \\
G_{\phi}^{\mu}(x,y,z) & \> = \> \bra{0}T\{j_A^{\mu}(x)\Phi(y)\Phi^{\dagger}(z)\}\ket{0}
\end{split}
\end{equation}
with the full propagators of the external legs cut off. The matrix $\Xi$,
\begin{equation}
\Xi=\left(\begin{array}{cr}
1 & 0\\
0 & -1
\end{array}\right),
\end{equation}
operates in the $\phi-\phi^{\dagger}$ space and is quite analogous to
$\gamma_5$ in the fermion sector.

\begin{figure}[t]
\begin{center}
\includegraphics{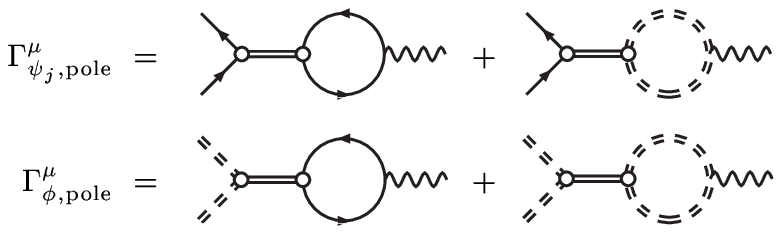}
\caption[]{The diagrammatical representation of pole parts of proper vertices
$\Gamma^{\mu}_{\psi_j}$ and $\Gamma^{\mu}_{\phi}$
(\ref{Ward_identities_pole}). The pole itself is interpreted as a propagator of
intermediate massless scalar particle --  the Nambu-Goldstone boson, depicted
by solid double line. The small empty circles are vertices $P_{\psi_j}$,
$P_{\phi}$ (\ref{ng_vertices}). The external vector boson lines are indicated
for the case the $U(1)_A$ symmetry was gauged.} \label{ng-pic}
\end{center}
\end{figure}

For $\Sigma_j$ and $\Pi$ different from zero these identities imply that the
proper vertices themselves develop poles for $q^{2}\rightarrow 0$:
\begin{equation}
\begin{split}
\Gamma^{\mu}_{\psi_j,\mathrm{pole}}(p+q,p) & \> = \> \frac{q^\mu}{q^2} Q_j \left[ S_j^{-1}(p+q)\gamma_5+\gamma_5S_j^{-1}(p) \right] \,,\\
\Gamma^{\mu}_{\phi,\mathrm{pole}}(p+q,p)   & \> = \> \frac{q^\mu}{q^2} \left[ -2D^{-1}(p+q)\Xi+2\Xi D^{-1}(p) \right] \,.
\end{split}
\label{Ward_identities_pole}
\end{equation}
In fact the poles are nothing but the manifestation of the presence of the NG
boson, see Fig. \ref{ng-pic}. These identities also allow to explicitly compute
the interaction vertices of the NG boson with the fermions and scalars; for the
details see Ref. \cite{Brauner:2005hw}, the result is [within the Ansatz
(\ref{ansatz})]
\begin{eqnarray}\label{ng_vertices}
  \lefteqn{ \!\!\!\!\!\!\!\!\hphantom{_{\phi}} P_{\psi_j}(p+q,p) = -\frac{Q_j}{N}\left[\Sigma_j(p+q)+\Sigma_j(p)\right]\gamma_5 \,,} \nonumber \\
    \lefteqn{ \!\!\!\!\!\!\!\!\hphantom{_{\psi_j}} P_{\phi}(p+q,p)  =  } \\
    && -\frac2N\left(
\begin{array}{cc}
0 & \Pi(p+q)+\Pi(p) \\
-\Pi^*(p+q)-\Pi^*(p) & 0
\end{array}\right) \nonumber    \,.
\end{eqnarray}
The normalization factor $N$ can be expressed as
\begin{equation}
N=\sqrt{J_{\psi_1}(0)+J_{\psi_2}(0)+J_{\phi}(0)} \,,
\end{equation}
where the functions $J_{\psi_j}$ and $J_{\phi}$ are defined by
\begin{widetext}
\begin{equation}
\begin{split}
-\I q^{\mu}J_{\psi_j}(q^2)&=8\int\frac{\d^4k}{(2\pi)^4}\frac{(k-q)^{\mu}\Sigma_{j,k}}{k^2-\Sigma^2_{j,k}}
\frac{\Sigma_{j,k}+\Sigma_{j,k-q}}{(k-q)^2-\Sigma^2_{j,k-q}} \,, \\
-\I q^{\mu}J_{\phi}(q^2)&=8\int\frac{\d^4k}{(2\pi)^4}\frac{(2k-q)^{\mu}(k^2-M^2)}{(k^2-M^2)^2-|\Pi_k|^2}
\frac{\Re\left[\Pi^*_{k-q}\bigl(\Pi_{k}+\Pi_{k-q}\bigr)\right]}{\bigl[(k-q)^2-M^2\bigr]^2-|\Pi_{k-q}|^2} \,.
\end{split}
\label{J_integrals}
\end{equation}
\end{widetext}

\section{Local Symmetry}

\subsection{Gauge boson mass}

\begin{figure}[t]
\begin{center}
\includegraphics[width=0.3\textwidth]{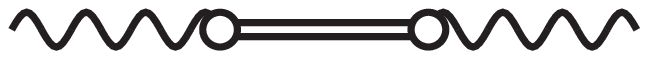}
\caption[]{The effective bilinear coupling of NG boson with gauge boson, giving
rise to the longitudinal polarisation of the latter and hence generating its
mass. The bilinear coupling is induced by the couplings of NG boson with
fermionic and scalar fields $P_{\psi_j}$ and $P_{\phi}$, compare with Fig.
\ref{ng-pic}.}
\label{bilinear}
\end{center}
\end{figure}

In this step we switch on the gauge interaction perturbatively and compute the
axial-vector gauge boson mass squared as a residue at the massless pole of the
gauge field polarization tensor $\Pi^{\mu \nu}(q)$. As in other strongly
coupled models in which the NG boson is a composite
\cite{Eidelman:2004wyTC,Freundlich:1970kn,Jackiw:1973tr,Cornwall:1973ts,Gribov:1994jy,Tanabashi:1989sz}
we are only able to compute the longitudinal part of $\Pi^{\mu \nu}$ due to the
bilinear coupling of NG boson with gauge boson -- see Fig. \ref{bilinear}.

Insisting on transversality of $\Pi^{\mu \nu}$ due to the axial-vector current
conservation, we conclude that the gauge boson mass squared is softly generated
in terms of $\Sigma_j$ and $\Pi$ as
\begin{equation}
    M_A^2 = g^2 [I_{\psi_1}(0)+I_{\psi_2}(0)+I_{\phi}(0)] \,,
\label{mass_vector}
\end{equation}
where
\begin{widetext}
\begin{equation}
\begin{split}
    \I q^\nu I_{\psi_j}(q^2)
    & =
    4 \int\frac{\d^4 k}{(2\pi)^4}
    \frac{\left[ (k+q)^\nu\Sigma_{j,k}-k^\nu\Sigma_{j,k+q} \right]
          \left[ \Sigma_{j,k+q}+\Sigma_{ji,k} \right]}
         {\Big[ (k+q)^2-\Sigma_{j,k+q}^2 \Big]
          \Big[ k^2-\Sigma_{j,k}^2 \Big]} \,,
    \\
    \I q^\nu I_{\phi}(q^2)
    & =
    -2 \int\frac{\d^4 k}{(2\pi)^4}
    \frac{ (2k+q)^\nu
          \Big\{ [\Pi_{k+q}+\Pi_{k}]
                \left[ [(k+q)^2-M^2]\Pi^*_{k}-(k^2-M^2)\Pi^*_{k+q}
                \right] + \cc
          \Big\} }
         { \Big[ [(k+q)^2-M^2]^2-|\Pi_{k+q}|^2 \Big]
           \Big[ (k^2-M^2)^2-|\Pi_{k}|^2 \Big] } \,.
\end{split}
\end{equation}
\end{widetext}
For $q^2=0$ we have explicitly
\begin{equation}
\begin{split}
    I_{\psi_j}(0)
    & =
    -8\I \int\frac{\d^4 k}{(2\pi)^4}
    \frac{\Sigma_{j,k} \left[ \Sigma_{j,k}-2k^2\frac{\d}{\d k^2} \Sigma_{j,k} \right]}
         {\Big[ k^2-\Sigma_{j,k}^2 \Big] ^2} \,,
    \\
    I_{\phi}(0)
    & =
    16\I \int\frac{\d^4 k}{(2\pi)^4}
    \frac{ k^2 |\Pi_{k}| \left[ |\Pi_{k}| - (k^2-M^2)\frac{\d}{\d k^2}|\Pi_{k}| \right] }
         { \Big[ (k^2-M^2)^2-|\Pi_{k}|^2 \Big] ^2 } \,.
\end{split}
\end{equation}

In the weakly coupled Higgs model with the elementary `would-be' NG boson the
spontaneous gauge boson mass generation is more transparent: The polarization
tensor $\Pi^{\mu \nu}(q)$ can easily be computed completely
\cite{Englert:1964et} and decent behavior of scattering amplitudes with
longitudinally polarized massive gauge bosons can be illustrated explicitly.

On the other hand, in strongly coupled models the lack of the bound-state
spectrum apart from the NG sector guaranteed by the existence theorem of
Goldstone implies: First, the $g^{\mu \nu}$ part of the polarization tensor can
hardly be computed. Second, decent behavior of scattering amplitudes with
longitudinally polarized massive gauge bosons cannot be determined from first
principles.

\subsection{$A^3$ vertex}

The spontaneous breakdown of the Abelian gauge chiral symmetry found in the
low-momentum behavior of propagators should by construction manifest in
noninvariant loop-generated vertices. Being ultraviolet-finite they represent
the genuine quantum-field theory predictions of the suggested nonperturbative
approach. In the Abelian prototype presented here it suffices to provide one
illustrative (still gedanken) example:

\begin{figure}[t]
\begin{center}
\includegraphics[width=0.4\textwidth]{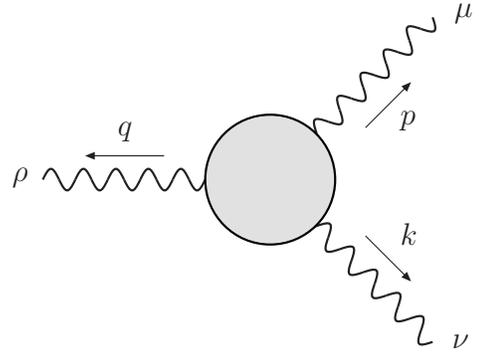}
\caption[]{The full $A^3$ vertex, denoted in the main text as $\I T^{\mu\nu\rho}(p,k)$. Energy-momentum conservation is assumed: $p+k+q=0$.}
\label{AAAfull}
\end{center}
\end{figure}

\begin{figure}[t]
\begin{center}
\includegraphics{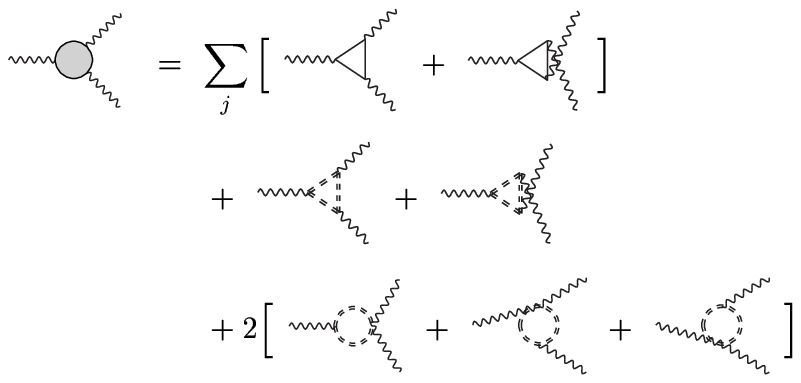}
\caption[]{Diagrammatical representation of the expression (\ref{AAAsum}). Each
diagram has the same assignment of momenta $p$, $k$, $q$ and Lorentz indices
$\mu$, $\nu$, $\rho$ as the full vertex in Fig. \ref{AAAfull}. The symmetry
factors are indicated explicitly.} \label{AAAsumPic}
\end{center}
\end{figure}

The gauge interaction of $A^{\mu}$ with the massive fermions gives rise to the
effective $A^3$ vertex depicted in Fig. \ref{AAAfull}. Up to $g^3$ the result
(calculated as a proper vertex) can be written as the following sum (compare
with Fig. \ref{AAAsumPic})
\begin{eqnarray}
    \lefteqn{ \I T^{\mu\nu\rho}(p,k)  =  } \nonumber \\
    && \sum_{j} \Big[ \I T^{\mu\nu\rho}_{\psi_j}(p,k) + \I T^{\nu\mu\rho}_{\psi_j}(k,p) \Big] \label{AAAsum} \\
    && {}   + \I T^{\mu\nu\rho}_\Phi(p,k) + \I T^{\nu\mu\rho}_\Phi(k,p) \nonumber \\
    && {} + 2 \Big[ \I T^{\mu\nu\rho}_4(p,k) + \I T^{\mu\rho\nu}_4(p,-p-k) + \I T^{\nu\rho\mu}_4(k,-p-k) \Big] \nonumber    \,,
\end{eqnarray}
where (with the $\I 0^+$'s suppressed)
\begin{widetext}
\begin{equation}
\begin{split}
    \I T^{\mu\nu\rho}_{\psi_j}(p,k)
    & \> = \>
    g^3 Q_j^3 \int\frac{\d^4 \ell}{(2\pi)^4}    \frac{1}{\left[\ell^2-\Sigma_{j,\ell}^2\right]\left[(\ell+p)^2-\Sigma_{j,\ell+p}^2\right]\left[(\ell-k)^2-\Sigma_{j,\ell-k}^2\right]} \times \\
    & \hphantom{ \> = \> }
    \Big\{  \Tr[\gamma^\mu\slashed{\ell}\gamma^\nu(\slashed{\ell}-\slashed{k})\gamma^\rho(\slashed{\ell}+\slashed{p})\gamma_5] +
    \\ &
    \hphantom{ \> = \> \{} - \, 4\I\epsilon^{\mu\nu\rho}_{\hphantom{\mu\nu\rho}\sigma}
    \left[      \ell^\sigma\Sigma_{j,\ell+p}\Sigma_{j,\ell-k}+(\ell+p)^\sigma\Sigma_{j,\ell}\Sigma_{j,\ell-k}+(\ell-k)^\sigma\Sigma_{j,\ell}\Sigma_{j,\ell+p}
    \right]
    \Big\} \,,
    \\
    \I T^{\mu\nu\rho}_\Phi(p,k)
    & \> = \>
    g^3 \!\!\int\!\frac{\d^4 \ell}{(2\pi)^4}
    \frac{ (2\ell+p)^\mu (2\ell-k)^\nu (2\ell+p-k)^\rho }
    {[ (\ell^2-M^2)^2 - |\Pi_\ell|^2 ]
     \big[ \left((\ell+p)^2-M^2\right)^2 - |\Pi_{\ell+p}|^2 \big]
     \big[ \left((\ell-k)^2-M^2\right)^2 - |\Pi_{\ell-k}|^2 \big]} \times
     \\ & \hphantom{ \> = \> }
     \Big\{ (\ell^2-M^2)^2\left[\Pi_{\ell+p}^*\Pi_{\ell-k}-\cc\right]+
     \\ &  \hphantom{ \> = \> \{} + \!
                     \left[(\ell+p)^2-M^2\right]^2\left[\Pi_{\ell-k}^*\Pi_{\ell}-\cc\right]+
                     \left[(\ell-k)^2-M^2\right]^2\left[\Pi_{\ell}^*\Pi_{\ell+p}-\cc\right]
     \Big\} \,,
     \\
    \I T^{\mu\nu\rho}_4(p,k)
    & \> = \>
    4 g^3 \int\frac{\d^4 \ell}{(2\pi)^4}
    \frac{g^{\mu\nu}(2\ell+p-k)^\rho\left[\Pi_{\ell+p}^*\Pi_{\ell-k}-\cc\right]}
    {\big[ \left((\ell+p)^2-M^2\right)^2 - |\Pi_{\ell+p}|^2 \big]
     \big[ \left((\ell-k)^2-M^2\right)^2 - |\Pi_{\ell-k}|^2 \big]} \,.
\end{split}
\label{amplitude_no_approx}
\end{equation}
\end{widetext}
Being a combination of three axial vectors, the amplitude should change its
sign under the operation of parity. Indeed it does: in the fermionic
contributions it is due to the presence of the Levi-Civita tensor, while in the
scalar contributions it is due to the following behavior of the scalar
self-energy under parity:
\begin{equation}
\Pi\stackrel{\mathcal{P}}{\rightarrow}\Pi^* \,.
\end{equation}

For illustration, let us now evaluate the amplitude $\I T^{\mu\nu\rho}(p,k)$
under certain approximations. First, we set the self-energies to be some
constants. Note that the integrals remain perfectly UV finite. In fact, the
scalar integrals vanish, while the quadratic divergences in the fermion
integrals cancel each other since
\begin{equation}
    \sum_j Q_j^3=0
\end{equation}
(this, together with the requirement $\sum_j Q_j=0$, is precisely the condition
for the axial anomaly to vanish). The linear divergences cancel because of the
symmetric integration. Note that within this approximation the result is
exactly the same as in the Higgs mechanism since we totally omit the nontrivial
structure of the propagators, represented by nonconstant self-energies, and
approximate them by the pole contribution. Second, for the sake of simplicity,
we put all external momenta on their mass-shell:
\begin{equation}
    p^2=k^2=q^2=M_A^2 \,.
\end{equation}
The energy-momentum conservation enables us to easily compute the dot products
of external momenta:
\begin{equation}
    p\cdot k=p\cdot q=k\cdot q=-\frac{1}{2}M_A^2 \,.
\end{equation}

Constant fermion self-energies turn out to be the masses:
$\Sigma_j(p^2)=\mathrm{const.}=m_j$. The resulting amplitude reads
\begin{equation}
    \I T^{\mu\nu\rho}(p,k)  =  G \left[
      p^\mu \epsilon^{\nu\rho}_{\hphantom{\nu\rho}\alpha\beta}
    - k^\nu \epsilon^{\mu\rho}_{\hphantom{\mu\rho}\alpha\beta}
    - (p+k)^\rho \epsilon^{\mu\nu}_{\hphantom{\mu\nu}\alpha\beta} \right] p^\alpha k^\beta \,,
\end{equation}
or equivalently
\begin{widetext}
\begin{equation}
    \I T^{\mu\nu\rho}(p,k)  = G \left[
    (q^\mu k^\alpha - k^\mu q^\alpha)p^\beta \epsilon^{\nu\rho}_{\hphantom{\nu\rho}\alpha\beta} +
    (p^\nu q^\alpha - q^\nu p^\alpha)k^\beta \epsilon^{\rho\mu}_{\hphantom{\rho\mu}\alpha\beta} +
    (k^\rho p^\alpha - p^\rho k^\alpha)q^\beta \epsilon^{\mu\nu}_{\hphantom{\mu\nu}\alpha\beta}
    \right] \,.
    \label{amplituda}
\end{equation}
\end{widetext}
The latter form is perhaps more convenient because the invariance under
exchange
\begin{equation}
    (p,\mu)\leftrightarrow(k,\nu)\leftrightarrow(q,\rho)
    \label{crossing_symmetry}
\end{equation}
is more apparent. The amplitude (\ref{amplituda}) can be generated via the
following effective Lagrangian:
\begin{equation}
    \eL_{\rm{eff}} = G \, \epsilon_{\alpha\beta\gamma\delta} (\partial_\sigma A^\alpha)(\partial^\beta A^\sigma)(\partial^\gamma A^\delta) \,.
\end{equation}

Note that in the result (\ref{amplituda}) there are no terms proportional to
$\epsilon^{\mu\nu\rho}_{\hphantom{\mu\nu\rho}\sigma}$, although they are
present in the original $\I T^{\mu\nu\rho}_{\psi_j}$ [see
(\ref{amplitude_no_approx})]. This is a simple consequence of the restriction
$p^2=k^2=q^2$, for in this case there is no way how to write down any
nontrivial linear combination of terms
$\epsilon^{\mu\nu\rho}_{\hphantom{\mu\nu\rho}\sigma}p^\sigma$,
$\epsilon^{\mu\nu\rho}_{\hphantom{\mu\nu\rho}\sigma}k^\sigma$ and
$\epsilon^{\mu\nu\rho}_{\hphantom{\mu\nu\rho}\sigma}q^\sigma$ which would be
invariant under the exchange (\ref{crossing_symmetry}). For general momenta
$p^2\neq k^2\neq q^2$ these terms would be present.

\begin{figure}[t]
\begin{center}
\framebox{\scalebox{0.8}{\includegraphics{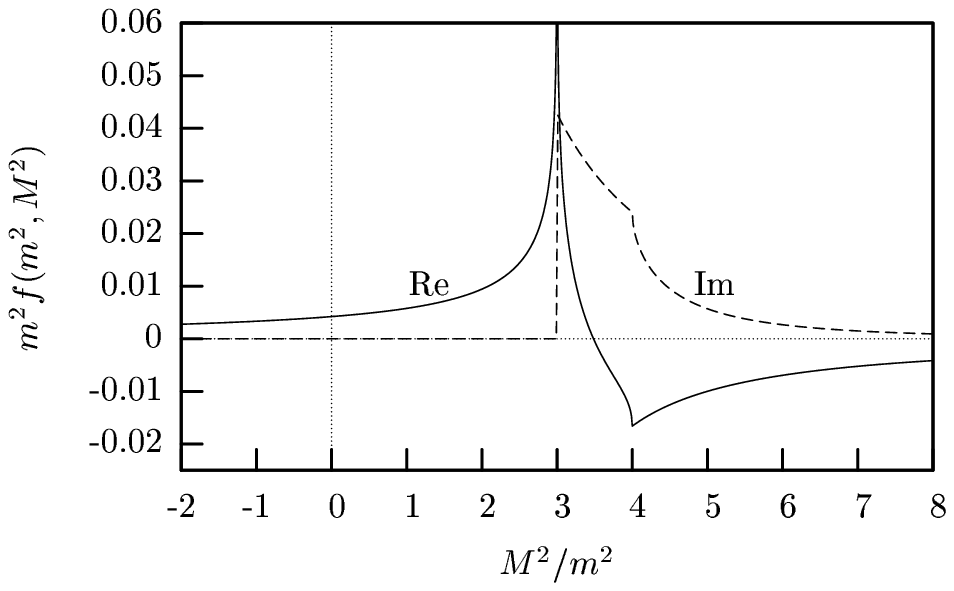}}}
\end{center}
\caption{The $M^2$-dependence of the function $f(m^2,M^2)$. Both quantities are
normalized by $m^2$ to be dimensionless; note that $m^2 f(m^2,M^2)$ is only
function of $M^2/m^2$. The cusps appear at $M^2=3m^2$ and $M^2=4m^2$. The
former one also indicates the beginning of the imaginary part.} \label{plot_g}
\end{figure}

The effective coupling constant $G$ can be expressed as
\begin{equation}
    G = g^3 \sum_j Q_j^3 f(m_j^2,M_A^2) \,.
\end{equation}
The function $f$ is defined by the integral
\begin{widetext}
\begin{equation}
    f(m^2,M^2) = \frac{2}{\pi^2 M^2}\int_0^1 \d z \frac{z(1-z)}{\sqrt{z(3z-4)+\frac{4m^2}{M^2}}}
    \arctan\frac{z}{\sqrt{z(3z-4)+\frac{4m^2}{M^2}}}
\end{equation}
\end{widetext}
(here $m^2$ should be replaced by $m^2-\I 0^+$ whenever the correct branch
choice of a multivalued analytic function is in question), which can be
calculated analytically in some special cases:
\begin{equation}
\begin{split}
    f(m^2,0) & \> = \> \frac{1}{24 \pi^2 m^2} \,, \\
    f(0,M^2) & \> = \> \frac{-1}{6 \pi^2 M^2} \,.
\end{split}
\end{equation}
More information about the shape of $f$ can be extracted numerically -- see
Fig. \ref{plot_g}.

\section{Numerics}
\label{numerics}

In this section the results of the numerical solution of Schwinger-Dyson
equations (\ref{SD_model_equations}) are presented. Moreover, as we consider
the results quite interesting and important, the description of the numerical
procedure together with discussion of its stability is presented.

\subsection{Numerical results}

The system of equations (\ref{SD_model_equations}) is quite difficult to be
solved even numerically, so certain approximations have to be done. First, we
switch to the Euclidean metric via the Wick rotation. By this we get rid of
some poles in the fermionic and scalar propagators. The absence of poles makes
numerical integration much easier. Moreover, one can consider the self-energies
to be real, without loss of generality. Note, however, that not all poles are
removed, namely the pole in the scalar propagator remains: It takes place if
$\Pi(p^2) = p^2+M^2$ for some $p^2$.

Second, we set $\lambda=0$. In fact, in the present model the $\lambda$-term is
not of high importance for the mass generation. In the Higgs case the scalar
self-interaction is of utmost importance: It fixes the numerical value of the
condensate and stabilizes the scalar field energy. In our case both reasons for
considering $\lambda \neq 0$ are clearly absent. However, in full treatment
including the perturbative effects of renormalization yet to be done the term
$\frac{1}{2}\lambda(\phi^\dag\phi)^2$ will become indispensable as a counter
term.

As a net result, we solve the following set of equations for the unknown
functions $\Sigma_1(p^2)$, $\Sigma_2(p^2)$ and $\Pi(p^2)$:
\begin{equation}
\begin{split}
    \Sigma_{1,p} & \> = \>
    y_1^2\int\frac{\d^4k}{(2\pi)^4}\frac{\Sigma_{1,k}}{k^2+\Sigma_{1,k}^2} \frac{\Pi_{k-p}}{\left[(k-p)^2+M^2\right]^2-\Pi_{k-p}^2} \,,
    \\
    \Sigma_{2,p} & \> = \>
    y_2^2\int\frac{\d^4k}{(2\pi)^4}\frac{\Sigma_{2,k}}{k^2+\Sigma_{2,k}^2}\frac{\Pi_{k-p}}{\left[(k-p)^2+M^2\right]^2-\Pi_{k-p}^2} \,,
    \\
    \Pi_p & \> = \>
    \sum_{j=1,2}2 y_j^2\int\frac{\d^4k}{(2\pi)^4}\frac{\Sigma_{j,k}}{k^2+\Sigma_{j,k}^2}\frac{\Sigma_{j,k-p}}{(k-p)^2+\Sigma_{j,k-p}^2}
    \,,
\end{split}
\label{sd_euclidean}
\end{equation}
where $p^2=p_0^2+p_1^2+p_2^2+p_3^2\geq 0$.

\begin{figure}[t]
\begin{center}
\framebox{\scalebox{0.8}{\includegraphics{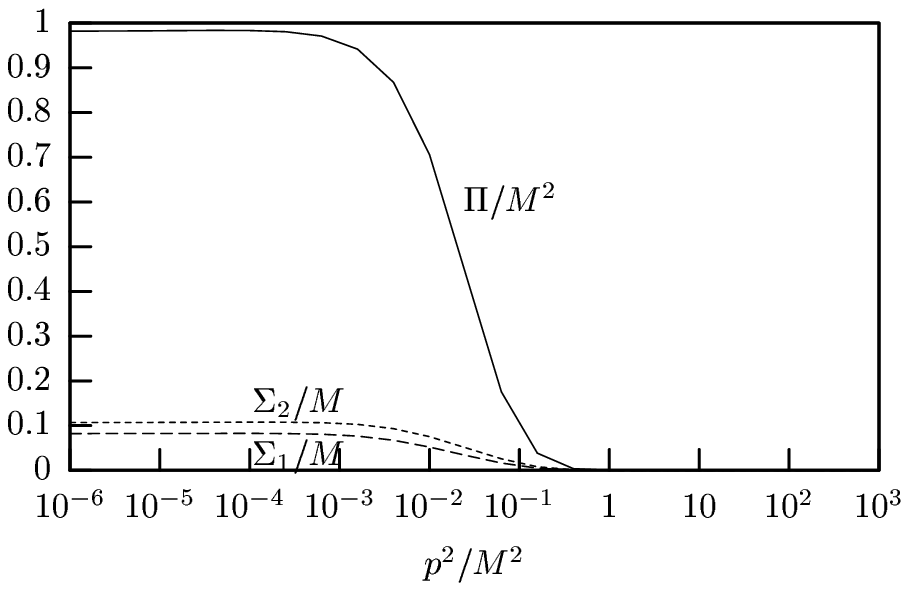}}}
\end{center}
\caption{Typical shape of the solutions $\Sigma_1(p^2)$, $\Sigma_2(p^2)$ and
$\Pi(p^2)$ to the system of equations (\ref{sd_euclidean}), computed here for
Yukawa coupling constants $y_1=83$ and $y_2=88$. Note the saturation of the
self-energies at low momenta and fast decrease at high momenta.}
\label{self_energies_pic}
\end{figure}

The solutions can be classified with respect to the pairs of coupling constants
$(y_1,y_2)$. The mass parameter $M$ is not relevant, since as the only
parameter with dimension of mass in the theory it serves just as a scale
parameter for the self-energies and momenta. The typical shape of the resulting
self-energies is depicted in Fig. \ref{self_energies_pic}. They are saturated
at low momenta and fall down rapidly at high momenta. Using logarithmic scale
on $y$-axis in Fig. \ref{self_energies_pic} it would be apparent that the
self-energies fall down actually faster then any power of $p^2$.

\begin{figure}[t]
\begin{center}
\scalebox{0.8}{\includegraphics{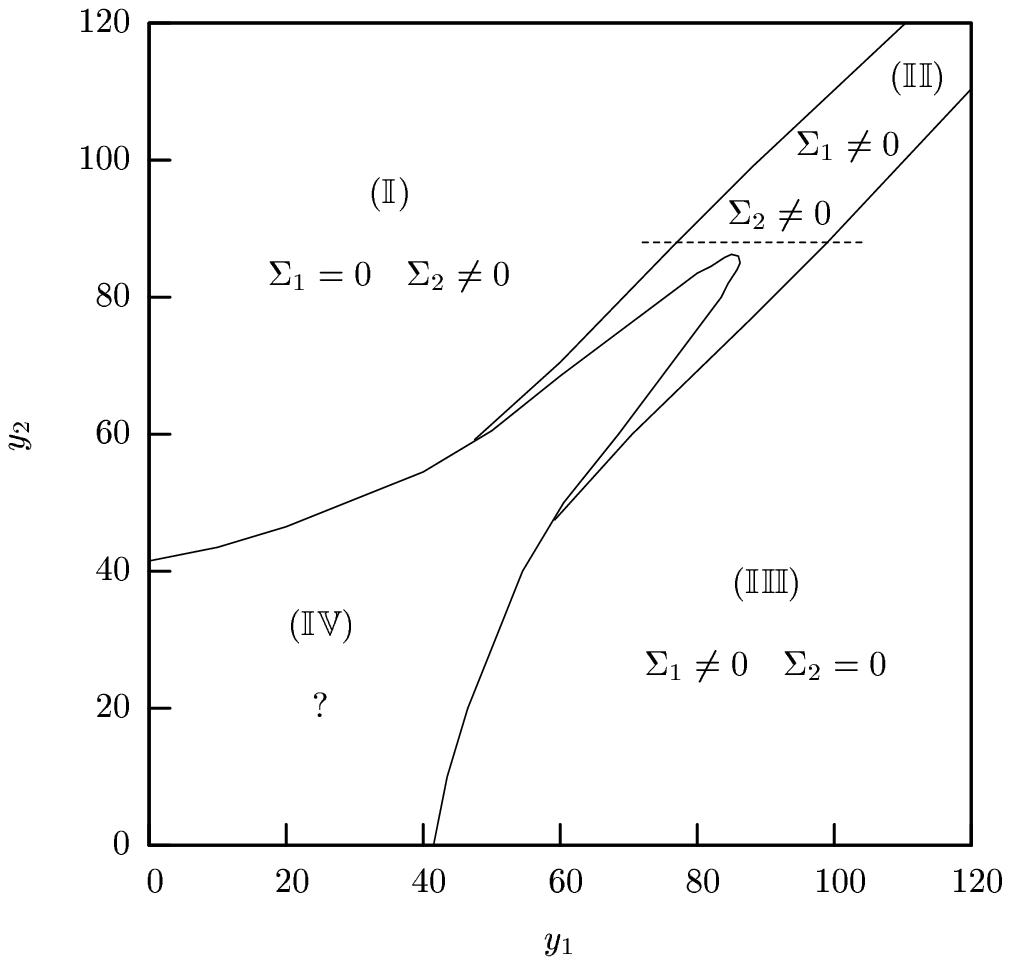}}
\end{center}
\caption{The $(y_1,y_2)$ plane with indicated areas of different behavior of
the system of equations (\ref{sd_euclidean}). According to the resulting
fermion self-energies there are three main areas: first where $\Sigma_1 = 0$
and $\Sigma_2 \neq 0$, second where $\Sigma_1 \neq 0$ and $\Sigma_2 \neq 0$ and
third where $\Sigma_1 \neq 0$ and $\Sigma_2 = 0$, denoted as $(\mathbb{I})$,
$(\mathbb{II})$ and $(\mathbb{III})$, respectively. There is also another
important area, denoted as $(\mathbb{IV})$, where the pole in the scalar
propagator takes place (and hence the imaginary parts of the self-energies are
expected to appear). However we do not know anything about behavior of the
self-energies in this area because of difficulties when numerically integrating
the pole. The dashed line, going from $y_1=72$ and $y_2=88$ to $y_1=104$ and
$y_2=88$, shows where the dependence of the spectrum on the Yukawa coupling
constants was probed -- see Fig. \ref{plot_Ms} and Fig. \ref{plot_Mf_MA}.}
\label{y_plane_pic}
\end{figure}

We have found interesting qualitative differences between the solutions of Eq.
(\ref{sd_euclidean}) for different values $(y_1,y_2)$. The main results are
summarised in Fig. \ref{y_plane_pic}. There is an area in the $(y_1,y_2)$
plane, located around the axis $y_1=y_2$ and denoted as $(\mathbb{II})$, where
both $\Sigma_1$ and $\Sigma_2$ are nonzero, and areas where only one of them,
$\Sigma_1$ or $\Sigma_2$, is nonzero [areas $(\mathbb{I})$ and
$(\mathbb{III})$]. There is also an area $(\mathbb{IV})$ where the pole in the
scalar propagator, mentioned earlier, matters. The self-energies (which are
expected to be imaginary here) were not computed in this area, we do not know
anything about their behavior here.

\begin{figure}[t]
\begin{center}
\framebox{\scalebox{0.8}{\includegraphics{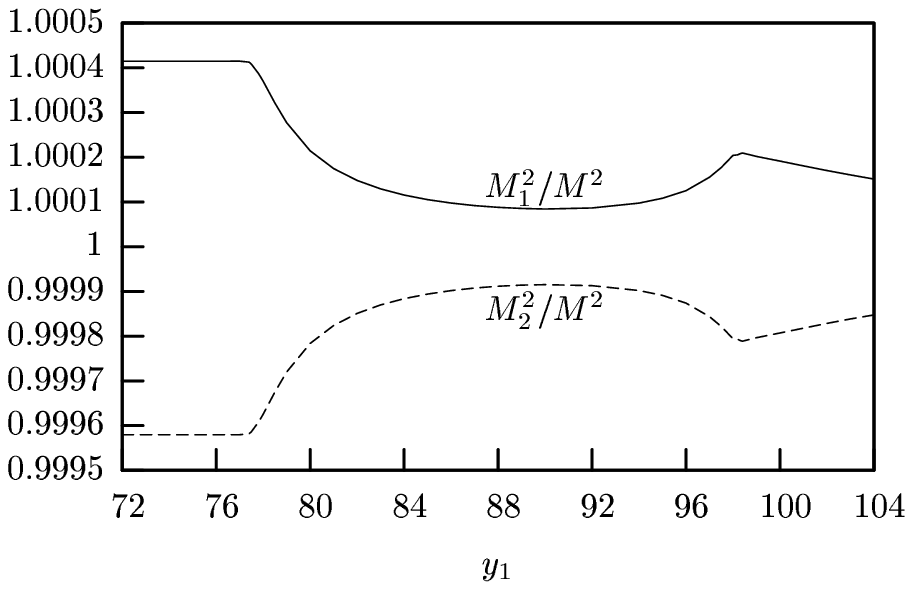}}}
\end{center}
\caption{The $y_1$-dependence of the scalar masses $M^2_{1,2}$ with fixed $y_2=88$.}
\label{plot_Ms}
\end{figure}

\begin{figure}[t]
\begin{center}
\framebox{\scalebox{0.8}{\includegraphics{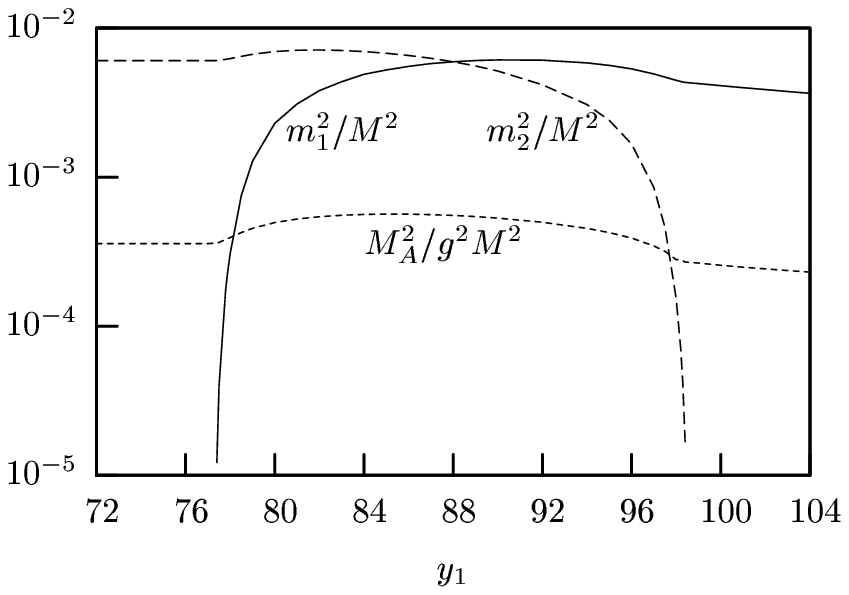}}}
\end{center}
\caption{The $y_1$-dependence of the fermion and vector boson masses
$m^2_{1,2}$ and $M^2_{A}$ with fixed $y_2=88$.}
\label{plot_Mf_MA}
\end{figure}

Our aim was to find the dependence of the spectrum -- the masses of the
fermions, scalar bosons and the vector boson -- on the Yukawa coupling
constants $y_1$ and $y_2$ [or, more precisely, on absolute values $|y_1|$ and
$|y_2|$, due to the shape of the Schwinger-Dyson equations
(\ref{sd_euclidean})]. For the calculation of masses we have used Eqs.
(\ref{mass_fermions}), (\ref{mass_scalars}), (\ref{mass_vector}), in the case
of vector boson Wick rotated to the Euclidean metric. We have probed the
$y_{1,2}$-dependence along the cut depicted in Fig. \ref{y_plane_pic}, since it
connects all the three main areas $(\mathbb{I})$, $(\mathbb{II})$ and
$(\mathbb{III})$ and therefore the resulting $y_{1,2}$-dependence of the
spectrum can be regarded as quite typical. The results are depicted in Fig.
\ref{plot_Ms} and Fig. \ref{plot_Mf_MA}. Note how the critical lines between
the areas are evident in the $y_{1,2}$-dependence of the spectrum. The most
significant result -- the behavior of the fermionic spectrum -- can be seen in
Fig. \ref{plot_Mf_MA}. As $y_1$ approaches the critical line between
$(\mathbb{II})$ and $(\mathbb{I})$ [or $(\mathbb{III})$, respectively] \emph{in
the direction} from $(\mathbb{II})$ to $(\mathbb{I})$ [$(\mathbb{III})$], the
ratio $m^2_2/m^2_1$ becomes arbitrarily high (low)!

\subsection{Numerical procedure}

Let us now say more about the numerical procedure itself. More formally, we
have to solve the following set of equations:
\begin{equation}
\begin{split}
    \Sigma_{1} & \> = \> F_1[\Sigma_{1},\Pi] \,, \\
    \Sigma_{2} & \> = \> F_2[\Sigma_{2},\Pi] \,, \\
    \Pi        & \> = \> G[\Sigma_{1},\Sigma_{2}] \,,
\end{split}
\end{equation}
where $F_j$, $G$ are some functionals. To solve this system, we have adopted
the method of iterations. After choosing some initial Ansatz for $\Sigma$'s:
\begin{equation}
    \Sigma_{1}^{(0)} , \Sigma_{2}^{(0)}
\end{equation}
and calculating also the `zeroth' iteration of the $\Pi$:
\begin{equation}
    \Pi^{(0)} = G[\Sigma_{1}^{(0)},\Sigma_{2}^{(0)}] \,,
\end{equation}
the iterating process is established: ($n \ge 1$)
\begin{equation}
\begin{split}
    \Sigma_{1}^{(n)} & \> = \> F_1[\Sigma_{1}^{(n-1)},\Pi^{(n-1)}] \,, \\
    \Sigma_{2}^{(n)} & \> = \> F_2[\Sigma_{2}^{(n-1)},\Pi^{(n-1)}] \,, \\
    \Pi^{(n)}        & \> = \> G[\Sigma_{1}^{(n-1)},\Sigma_{2}^{(n-1)}] \,.
\end{split}
\end{equation}
If this procedure converges, we have the solution.
In order to control the convergence we used the quantity
\begin{equation}
    I^{(n)}  =  \frac{\int X^{(n)}}{\int X^{(n-1)}}  \,,
\end{equation}
where $X$ stands for any of $\Sigma_j$, $\Pi$. If $I^{(n)}\to 1$, we considered
$X^{(n)}$ to be the solution. This simple criterion is sufficient in our case
because the functions $X^{(n)}$ turn out to be well behaved: there are no
intersections between distinct iterations, or, loosely speaking, their shapes
are similar, the only changes are in their `size' (at least in the
nonasymptotic region).

Usual behavior of such a nonlinear system in the case of only \emph{one
equation for one unknown function} (for instance an equation for $\Sigma$ with
$\Pi$ set to be a constant) is such that for \emph{any} initial Ansatz the
iteration procedure converges to some solution. The solution may be trivial
(i.e. $I^{(n)} < 1$, $I^{(n)} \nrightarrow 1$) or nontrivial (i.e. $I^{(n)} \to
1$), depending whether $y < y_\mathrm{crit}$ or $y > y_\mathrm{crit}$,
respectively, for some critical value $y_\mathrm{crit}$.

In our case of more \emph{coupled} equations the situation is, however,
different. Having some fixed Ansatz, there are three possibilities: (1) The
iteration procedure converges to the trivial solution [which exists always for
all $(y_1,y_2)$]. (2) It blows up (i.e. $I^{(n)} > 1$, $I^{(n)} \nrightarrow
1$). (3) It converges to a nontrivial solution (if it exists), depending on the
$(y_1,y_2)$ (see Fig. \ref{y_plane_pic}).

This behavior clearly depends on the Ansatz. To be specific, our class of
Ans\"{a}tze consisted of functions
\begin{equation}
    \Sigma_{1}^{(0)}(p^2) = \Sigma_{2}^{(0)}(p^2) = x \frac{M^5}{(p^2+M^2)^2}
    \label{ansatz_num}
\end{equation}
with $x$ being a free parameter. For $x$ too small or too large the iteration
procedure goes to the trivial solution or blows up, respectively. Properly
adjusting the value of $x$ between these two extremes one can manage that the
iteration procedure converges to the nontrivial solution.

\subsection{Numerical stability}

A special care was taken to check whether the numerical procedure is stable in
the sense that its results (especially the strong $y_{1,2}$-dependence of the
fermion masses) remain unchanged upon the changes of the numerical algorithm.
We tested four main variations of the algorithm:
\begin{description}
    \item[Class of Ans\"{a}tze] Besides (\ref{ansatz_num}) we considered also other Ansatz classes, all functions decreasing as some power of $p^2$ or exponentially, respectively. For some of them (some very rapidly decreasing exponentials) it was not possible to adjust the variable $x$ [see (\ref{ansatz_num})] to converge to any but the trivial solution. For the other Ansatz classes, however, the results were nontrivial and they all coincided.
    \item[Step size] There is, of course, a step size dependence of the numerical integration results, the important question is, however, how this dependence behaves for arbitrarily small step sizes. If there is no sensible (i.e. finite) limit of the integral as the step sizes are going to zero (the continuum limit), the results of the numerical integration have no meaning. We checked that this limit does exist and that all described phenomena, especially the strong $y_{1,2}$-dependence of the fermion masses, are present in it.
    \item[Integration method] Having expressed the integral equations (\ref{sd_euclidean}) in the hyperspherical coordinates, we can do two angular integrations analytically. The two remaining integrations (one over momentum and one angular) must be performed numerically. For this purpose we employed consecutively the trapezoidal rule and the Simpson's rule, for the angular integration also the Gauss-Chebyshev quadrature formula (using Chebyshev polynomials of the second kind). The final results for all integration methods agreed, the differences were only in the step size dependence, i.e. in the speed of convergence to the continuum limit.
    \item[Momentum cut-off] Since the momentum integral is over the infinite interval, for the purposes of the numerical integration a momentum cut-off must be introduced. However it turns out that one does not need to check the cut-off dependence of the results. That is because the resulting self-energies are so rapidly decreasing that the contribution from the high momenta is negligible.
\end{description}

Moreover, in order to check the consistency of our numerical method by a
comparison with an independent result, we calculated the equation for
$\Sigma_j$ [see (\ref{sd_euclidean})] with $\Pi$ set to be a constant [up to
our knowledge, there are no independent calculations of the full set of the
coupled equations (\ref{sd_euclidean}) we could compare with] and compared our
result with the results of \cite{Sauli:2006ba} [Eq. (2.11) and Fig. 3 therein].
They coincided.

\section{Concluding remarks and outlook}

Within the effective quantum field theory for electroweak interactions at
forthcoming energy scale we find the use of scalar fields still welcome: Their
Yukawa couplings with massless elementary fermion fields break explicitly the
huge chiral symmetries of three electroweakly identical fermion families
exactly to the sanctified $SU(2)_L\times U(1)_Y$. Alternatives without scalars
struggle with guaranteeing unobservability of physical consequences of these
unwanted symmetries.

Linear dependences of the wildly wide, sparse and irregular fermion mass
spectrum upon the Yukawa coupling constants of the Standard model are, however,
a drawback. It is feasible that the fermion masses can be generated by the
Yukawa interactions dynamically without ever referring to the scalar field
condensation \cite{Imri:1970,Tanabashi:1989sz}. In such a case the dependences
of the fermion masses upon the Yukawa coupling constants would be nonanalytic.
In an Abelian prototype we have provided a bona fide indication that this is
possible. Most importantly, we have convincingly demonstrated that a small
change in the ratio of the Yukawa couplings may lead to an orders-of-magnitude
change in the ratio of the generated fermion masses. We hope that this
mechanism could provide a natural explanation of the hierarchy of fermion
masses in the Standard model.

The price we pay is that the Yukawa interactions having the expected properties
must be strong. But we are accustomed to live with QCD strongly interacting at
large distances even though we still don't know how to solve it (incidentally,
apart from the NG sector). We are tempted to interpret our findings as
manifestations of nontrivial fixed points \cite{Wilson:1971ag} in our model.

The bonus we get is an interesting relation of fermion and gauge boson mass
generation mechanisms. Indeed, in contrast with the Higgs mechanism in our case
for the Yukawa couplings set to zero both fermion and gauge boson masses
vanish.

We are confident that with some technical modifications the generalization of
the Abelian mechanism presented above to $SU(2)_L \times U(1)_Y$ can be done
\cite{Brauner:2004kg}. Whether it can be converted into a phenomenologically
viable alternative of the Standard model remains, however, to be seen.

\begin{acknowledgments}
One of the authors (P.B.) is grateful to IPNP, Charles University in Prague,
and the ECT* Trento for the support during the ECT* Doctoral Training Programme
2005. This work was supported in part by the Institutional Research Plan
AV0Z10480505, by the GACR doctoral project No. 202/05/H003 and by the GACR
grant No. 202/06/0734.
\end{acknowledgments}


\end{document}